\documentstyle[aps,twocolumn]{revtex}
\begin{document}
\author{P. Salgado$^{1\thanks{
pasalgad@udec.cl}},$ M$.$ Cataldo$^{2\thanks{
mcataldo@ubiobio.cl}}$ and S. del Campo$^{3\thanks{
sdelcamp@ucv.cl}}$}
\address{$^1$Departamento de F\'{\i}sica, Universidad de Concepci\'{o}n,\\
Casilla 160-C, Concepci\'{o}n, Chile.\\
$^2$Departamento de F\'{\i}sica, Universidad del B\'{\i}o-B\'{\i}o,\\
Casilla 5-C, Concepci\'{o}n, Chile.\\
$^3$Instituto de F\'{i}sica, Universidad Cat\'{o}lica de Valpara\'{i}so,\\
Avda Brasil 2950, Valpara\'{i}so, Chile.}
\title{{\bf Higher dimensional gravity invariant under the Poincar\`e group}}
\maketitle

\begin{abstract}
It is shown that the Stelle-West Grignani-Nardelli-formalism allows, both
when odd dimensions and when even dimensions are considered, constructing
actions for higher dimensional gravity invariant under local Lorentz
rotations and under local Poincar\`{e} translations. It is also proved that
such actions have the same coefficients as those obtained by Troncoso and
Zanelli in ref.\cite{tron}.
\end{abstract}

\section{\bf Introduction}

The most general action for the metric satisfying the criteria of general
covariance and second-order field equations for $d>4$ is a polynomial of
degree $\left[ d/2\right] $ in the curvature known as the Lanczos-Lovelock
gravity theory $\left( LL\right) $ \cite{lanc},\cite{lovel}. The $LL$
lagrangian in a $d$-dimensional Riemannian manifold can be defined as a
linear combination of the dimensional continuation of all the Euler classes
of dimension $2p<d$ \cite{zum},\cite{teit}:
\begin{equation}
S=\int \sum_{p=0}^{\left[ d/2\right] }\alpha _pL^{(p)}  \label{uno}
\end{equation}
where $\alpha _p$ are arbitrary constants and
\begin{equation}
L_p=\varepsilon _{a_1a_2\cdot \cdot \cdot \cdot \cdot \cdot
A_d}R^{a_1a_2}\cdot \cdot \cdot \cdot R^{a_{2p-1}a_{2p}}e^{a_{2p+1}}\cdot
\cdot \cdot \cdot e^{a_d}  \label{dos}
\end{equation}
with $R^{ab}=d\omega ^{ab}+\omega _c^a\omega ^{cb}.$ The expression (\ref
{uno}) can be used both for even and for odd dimensions.

The large number of dimensionful constants in the $LL$ theory
$\alpha _p,$ $ p=0,1,\cdot \cdot \cdot ,\left[ d/2\right] ,$ which
are not fixed from first principles, contrast with the two
constants of the Einstein-Hilbert action.

In ref. \cite{tron} it was found that these parameters can be fixed in terms
of the gravitational and the cosmological constants, and that the action in
odd dimensions can be formulated as a gauge theory of the $AdS$ group, and
in a particular case, as a gauge theory of the Poincar\'{e} group. This
means that the action is invariant not only under standard local Lorentz
rotations $\delta e^a=\kappa _b^ae^b;$ $\delta \omega ^{ab}=-D\kappa ^{ab},$
but also under local $AdS$ boost $\delta e^a=D\rho ^a$ ; $\delta \omega
^{ab}=\frac 1{l^2}(\rho ^ae^b-\rho ^be^a),$ where $l$ is a length parameter.
They also show that this situation is not possible in even dimensions where
the action is invariant only under local Lorentz rotations in the same way
as is the Einstein-Hilbert action.

On the other hand, in ref. \cite{grigna} it was found that, in four
dimensions, gravity can be formulated as a gauge theory of the Poincar\`{e}
group. In this reference was considered the Poincar\`{e} gauge theory as
closely as possible to any ordinary non-Abelian gauge theory.

The basic idea of this formulation is founded on the mathematical
definition \cite{kobaya},\cite{stelle} of the vierbein $V^a.$ This
vierbein, also called solder form \cite{witten} was considered as
a smooth map between the tangent space to the space-time manifold
$M$ at a point $P$ with coordinates $x^\mu ,$ and the tangent
space to the internal $AdS$ space at the point whose $AdS$
coordinates are $\xi ^a(x)$, as the point $P$ ranges over the
whole manifold $M.$ The $fig.1$ of ref. \cite{stelle} illustrates
that such a vierbein $V_\mu ^a(x)$ is the matrix of the map
betweeen the tangent space $T_x(M)$ to the space-time manifold at
$x^\mu ,$ and the tangent space $ T_{\xi (x)}\left( \left\{
G/H\right\} _x\right) $ to the internal $AdS$ space $\left\{
G/H\right\} _x$ at the point $\xi ^a(x),$ whose explicit form is
given by $eq.(3.19)$ of ref. \cite{stelle}$.$

Taking the limit $m\rightarrow 0$ in such $eq.(3.19)$ of ref. \cite{stelle}
we obtain $V_\mu ^a(x)=D_\mu \xi ^a+e_\mu ^a$ which is the map between the
tangent space $T_x(M)$ to the space-time manifold at $x^\mu $ and the
tangent space $T_{\xi (x)}\left( \left\{ ISO(3,1)/SO(3,1)\right\} _x\right) $
to the internal Poincar\'e space $\left\{ ISO(3,1)/SO(3,1)\right\} _x$ at
the point $\xi ^a(x)$. The same result was obtained in ref. \cite{grigna} by
gauging the action of a free particle defined in the internal Minkowski
space. Here $\xi ^a$, called Poincar\`e coordinates, transform as a vector
under Poincar\`e transformations and can be interpreted as coordinates of an
internal Minkowskian space $M_\xi $ that can locally be made to coincide
with the tangent space. Any choice of Poincar\'e coordinates is equivalent
to a gauge that leaves the theory invariant under residual local Lorentz
transformations.

It is the purpose of this article to show that it is also possible to find
an action for higher dimensional gravity genuinely invariant under the
Poincar\'{e} group provided that one chooses the vierbein in an appropriate
way.

\section{\bf The Lanczos-Lovelock Gravity Theory}

In this section we shall review some aspects of higher dimensional gravity.
The main point of this section is to display the differences between the
invariances of $LL$ action when odd and even dimensions are considered.

\subsection{\bf The local AdS Chern-Simons and Born-Infeld like gravity}

The $LL$ action is a polynomial of degree $\left[ d/2\right] $ in curvature,
which can be written in terms of the Riemann curvature and the vielbein $e^a$
in the form (\ref{uno}), (\ref{dos}). In first order formalism the $LL$
action is regarded as a functional of the vielbein and spin connection, and
the corresponding field equations obtained by varying with respect to $e^a$
and $\omega ^{ab}$ read \cite{tron}:

\begin{equation}
\varepsilon _a=\sum_{p=0}^{\left[ \left( d-1\right) /2\right] }\alpha
_p(d-2p)\varepsilon _a^p=0  \label{tres}
\end{equation}
\begin{equation}
\varepsilon _{ab}=\sum_{p=1}^{\left[ \left( d-1\right) /2\right] }\alpha
_pp(d-2p)\varepsilon _{ab}^p=0  \label{cuatro}
\end{equation}
where we have defined
\[
\varepsilon _a^p:=\varepsilon _{ab_1b_2\cdot \cdot \cdot \cdot \cdot \cdot
b_{d-1}}R^{b_1b_2}\cdot \cdot \cdot
\]
\begin{equation}
\cdot \cdot \cdot R^{b_{2p-1}b_{2p}}e^{b_{2p+1}}\cdot \cdot \cdot \cdot
e^{b_{d-1}}  \label{cinco}
\end{equation}
\[
\varepsilon _{ab}^p=\varepsilon _{aba_3\cdot \cdot \cdot \cdot \cdot \cdot
a_d}R^{a_3a_4}\cdot \cdot \cdot
\]
\begin{equation}
\cdot \cdot \cdot \cdot R^{a_{2p-1}a_{2p}}T^{a_{2p+1}}e^{a_{2p+2}}\cdot
\cdot \cdot \cdot e^{a_d}.  \label{seis}
\end{equation}
Here $T^a=de^a+\omega _b^ae^b$ is the torsion $2$-form. Using the Bianchi
identity one finds \cite{tron}
\begin{equation}
D\varepsilon _a=\sum_{p=1}^{\left[ \left( d-1\right) /2\right] }\alpha
_{p-1}(d-2p+2)(d-2p+1)e^b\varepsilon _{ba}^p.  \label{siete}
\end{equation}
Moreover
\begin{equation}
e^b\varepsilon _{ba}=\sum_{p=1}^{\left[ \left( d-1\right) /2\right] }\alpha
_pp(d-2p)e^b\varepsilon _{ba}^p.  \label{ocho}
\end{equation}

 >From (\ref{siete}) and (\ref{ocho}) one finds for $d=2n-1$
\[
\alpha _p=\alpha _0\frac{(2n-1)(2\gamma )^p}{(2n-2p-1)}\left( {n-1
\atop p} \right) ;\qquad \alpha _0=\frac \kappa {dl^{d-1}};\qquad
\]
\begin{equation}
\gamma =-sign(\Lambda )\frac{l^2}2,  \label{nueve}
\end{equation}
where for any dimensions $l$ is a length parameter related to the
cosmological constant by $\Lambda =\pm (d-1)(d-2)/2l^2$

With these coefficients, the $LL$ action is invariant under local Lorentz
rotations and under local AdS boosts.

For $d=2n$ it is necessary to write equation (\ref{siete}) in the form \cite
{tron}
\[
D\varepsilon _a=T^a\sum_{p=1}^{\left[ n-1\right] }2\alpha
_{p-1}(n-p+1){\cal T}_{ab}^p
\]
\begin{equation}
-\sum_{p=1}^{\left[ n-1\right] }4\alpha _{p-1}(n-p+1)(n-p)e^b\varepsilon
_{ba}^p  \label{diez}
\end{equation}
with
\begin{equation}
{\cal T}_{ab}=\frac{\delta L}{\delta R^{ab}}=\sum_{p=1}^{\left[ \left(
d-1\right) /2\right] }\alpha _pp{\cal T}_{ab}^p  \label{once}
\end{equation}
where
\[
{\cal T}_{ab}^p=\varepsilon _{aba_3\cdot \cdot \cdot \cdot
a_d}R^{a_3a_4}\cdot \cdot \cdot
\]
\begin{equation}
\cdot \cdot \cdot R^{a_{2p-1}a_{2p}}T^{a_{2p+1}}e^{a_{2p+2}}\cdot \cdot
\cdot e^{a_d}.  \label{doce}
\end{equation}

The comparison between (\ref{ocho}) and (\ref{diez}) leads to \cite{tron}
\begin{equation}
\alpha _p=\alpha _0(2\gamma )^p {n \choose p}.  \label{trece}
\end{equation}

With these coefficients the $LL$ action, in the same way as the Hilbert
action, is invariant only under local Lorentz rotations.

\subsection{\bf Theories described by a generalized action}

Recently a class of gravitation theories was found \cite{criso} described by
the action
\begin{equation}
S=\int \sum \alpha _pL^{(p)}=\kappa \int \sum_{p=0}^kC_p^kL^{(p)}  \label{d1}
\end{equation}
where
\begin{equation}
\alpha _p=\kappa C_p^k=\{ {\textstyle {\kappa \frac{\alpha
_0d(2\gamma )^p}{(d-2p)}{k \choose p},\quad p\leq k \atop 0\qquad
p>k}} \label{d1'}
\end{equation}
with $1\leq k\acute \leq [(d-1)/2]$ and where $L_p$ is given by
\begin{equation}
L_p=\varepsilon _{a_1a_2\cdot \cdot \cdot \cdot \cdot \cdot
a_d}R^{a_1a_2}\cdot \cdot \cdot \cdot R^{a_{2p-1}a_{2p}}e^{a_{2p+1}}\cdot
\cdot \cdot \cdot e^{a_d}  \label{d2}
\end{equation}
where $R^{ab}=d\omega ^{ab}+\omega _c^a\omega ^{cb}$ is the
curvarure $2$ -form and $e^a$ is the vielbein $1$-form. For a
given dimension $d$, the $ \alpha _p$ coefficients give rise to a
family of inequivalent theories, labeled by the integer $k\in
\left\{ 1\cdot \cdot \cdot \cdot [(d-1)/2]\right\} $ which
represents the highest power of curvature in the lagrangian. This
set of theories possees only two fundamental constants, $ \kappa $
and $l$, related to the gravitational constant $G$ and the
cosmological constant $\Lambda $ \cite{tron}. For $k=1$, the
Einstein-Hilbert action is recovered, while for the largest value
of $k$, that is $k=[(d-1)/2],$ Born-Infeld and Chern-Simons
theories are obtained. These three cases exhaust the different
possibilities up to six dimensions, and new interesting cases
arise for $d\geq 7$. For instance, the case with $ k=2$, which is
described by the action \cite{criso}
\[
S_2=\kappa \int \varepsilon _{a_1a_2\cdot \cdot \cdot \cdot
a_d}(\frac{l^{-4} }de^{a_1}\cdot \cdot \cdot
e^{a_d}+\frac{2l^{-2}}{d-2}R^{a_1a_2}e^{a_3}\cdot \cdot \cdot
e^{a_d}
\]
\begin{equation}
+\frac 1{d-4}R^{a_1a_2}R^{a_3a_4}e^{a_5}\cdot \cdot \cdot e^{a_d})
\end{equation}
exists only for $d>4:$ in five dimensions this theory is
equivalent to Chern-Simons; for $d=6$ it is equivalent to
Born-Infeld and for $d=7$ and $ up $, if defines a new class of
theories.

At the end of the range, $k=[(d-1)/2]$, even and odd dimensions must be
distinguished. When $d=2n-1$, the maximum value of $k$ is $n-1$, and the
corresponding lagrangian of the action
\[
S=\int \sum_{p=0}^{\left[ d/2\right] }\frac \kappa {d-2p} {n-1
\choose p} l^{2p-d+1}\varepsilon _{a_1a_2\cdot \cdot \cdot \cdot
\cdot \cdot a_d}R^{a_1a_2}\cdot \cdot \cdot
\]
\begin{equation}
\cdot \cdot \cdot \cdot R^{a_{2p-1}a_{2p}}e^{a_{2p+1}}\cdot \cdot \cdot
\cdot e^{a_d}
\end{equation}
is a Chern-Simons $\left( 2n-1\right) $-form. This action \cite{tron} is
invariant not only under standard local Lorentz rotations
\begin{equation}
\delta e^a=\kappa _{\text{ }b}^ae^b,\quad \delta \omega ^{ab}=-D\kappa ^{ab},
\label{d3}
\end{equation}
but also under a local $AdS$ boost
\begin{equation}
\delta e^a=D\rho ^a,\quad \delta \omega ^{ab}=\frac 1{l^2}\left( \rho
^ae^b-\rho ^be^a\right)  \label{d4}
\end{equation}

For $d=2n$ and $k=n-1$, the action is given by
\[
S=\int \sum_{p=0}^{\left[ d/2\right] }\frac \kappa {2n} {n \choose
p} l^{2p-d+1}\varepsilon _{a_1a_2\cdot \cdot \cdot \cdot \cdot
\cdot a_d}R^{a_1a_2}\cdot \cdot \cdot
\]
\begin{equation}
\cdot \cdot \cdot R^{a_{2p-1}a_{2p}}e^{a_{2p+1}}\cdot \cdot \cdot \cdot
e^{a_d}.
\end{equation}
This action \cite{tron} is invariant under standard local Lorentz rotations
but it is not invariant under local $AdS$ boosts.

\section{\bf Lanczos-Lovelock Theory and the Poincar\`e group}

In this section we shall review some aspects of the
Stelle-West-Grignani-Nardelli $(SWGN)$-formalism. This formalism leads to a
formulation of general relativity where the Hilbert action is invariant both
under local Poincar\`{e} translations and under local Local Lorentz
rotations. The main point of this section is to show that the $\left(
SWGN\right) $ formalism permits, both when odd dimensions and when even
dimensions are considered, constructing a higher dimensional gravity action
invariant under local Lorentz rotations and under local Poincare
translations. It is also proved that such an action has the same
coefficients as those obtained in ref.\cite{tron}.

\subsection{\bf Einstein gravity invariant under the Poincar\`e group}

\subsubsection{{\bf Invariance of} {\bf the Hilbert action}}

The generators of the Poincar\`e group $P_a$ and $J_{ab\text{ }}$satisfy the
Lie algebra,
\[
\left[ P_a,P_b\right] =0;
\]
\[
\left[ J_{ab},P_c\right] =\eta _{ac}P_b-\eta _{bc}P_a;
\]
\begin{equation}
\left[ J_{ab},J_{cd}\right] =\eta _{ac}J_{bc}-\eta _{bc}J_{ad}+\eta
_{bd}J_{ac}-\eta _{ad}J_{bc}.  \label{do}
\end{equation}
Here the operators carry Lorentz indices not related to coordinate
transformations. The Yang-Mills connection for this group is given by
\begin{equation}
A=A^aT_a=e^aP_a+\frac 12\omega ^{ab}J_{ab}.  \label{tre}
\end{equation}
Using the algebra (\ref{dos}) and the general form for the gauge
transformations on $A$
\begin{equation}
\delta A=\nabla \lambda =d\lambda +\left[ A,\lambda \right]  \label{cuatr}
\end{equation}
with
\begin{equation}
\lambda =\rho ^aP_a+\frac 12\kappa ^{ab}J_{ab},  \label{cinc}
\end{equation}
we obtain that $e^a$ and $\omega ^{ab},$ under Poincar\`e translations,
transform as
\begin{equation}
\delta e^a=D\rho ^a;\quad \delta \omega ^{ab}=0,  \label{sei}
\end{equation}
and under Lorentz rotations, as
\begin{equation}
\delta e^a=\kappa _b^ae^b;\quad \delta \omega ^{ab}=-D\kappa ^{ab},
\label{siet}
\end{equation}
where $D$ is the covariant derivative in the spin connection $\omega ^{ab}$.
The corresponding curvature is
\[
F=F^aT_a=dA+AA
\]
\begin{equation}
=T^aP_a+\frac 12R^{ab}J_{ab}  \label{och}
\end{equation}
where
\begin{equation}
T^a=De^a=de^a+\omega _b^ae^b  \label{nuev}
\end{equation}
is the torsion $1$-form, and
\begin{equation}
R^{ab}=d\omega ^{ab}+\omega _c^a\omega ^{cb}  \label{die}
\end{equation}
is the curvature $2$-form.

The Hilbert action
\begin{equation}
S_{EH}=\int \varepsilon _{abcd}R^{ab}e^ce^d,  \label{onc}
\end{equation}
is invariant under diffeomorphism and under Lorentz rotations, but is not
invariant under Poincar\`e translations. In fact
\[
\delta S_{EH}=2\int \varepsilon _{abcd}R^{ab}e^c\delta e^d
\]
\begin{equation}
=2\int \varepsilon _{abcd}R^{ab}T^c\rho ^d  \label{doc}
\end{equation}
where we see that the invariance of the action requires imposing the torsion
free condition
\begin{equation}
T^a=De^a=de^a+\omega _b^ae^b=0.  \label{trec}
\end{equation}

\subsubsection{\bf The Stelle-West Grignani-Nardelli formalism}

The key ingredients of the $\left( SWGN\right) $ formalism are the so called
Poincare coordinates $\xi ^a(x)$ which behave as vectors under $ISO(3,1)$
and are involved in the definition of the $1$-form vierbein $V^a$, which is
not identified with the component $e^a$ of the gauge potential, but is given
by
\begin{equation}
V^a=D\xi ^a+e^a=d\xi ^a+\omega _b^a\xi ^b+e^a.  \label{catorce}
\end{equation}

Since $\xi ^a,$ $e^a,$ $\omega ^{ab}$ under local Poincare translations
change as
\begin{equation}
\delta \xi ^a=-\rho ^a,\quad \delta e^a=D\rho ^a,\quad \delta \omega ^{ab}=0;
\label{a}
\end{equation}
and under local Lorentz rotations change as
\begin{equation}
\delta \xi ^a=\kappa _{\text{ }b}^a\xi ^b,\quad \delta e^a=\kappa
_{\text{ } b}^ae^b,\quad \delta \omega ^{ab}=-D\kappa ^{ab};
\label{b}
\end{equation}
we have that the vierbein $V^a$ is invariant under local Poincare
translations
\begin{equation}
\delta V^b=0  \label{poinc}
\end{equation}
and, under local Lorentz rotations, transforms as
\begin{equation}
\delta V^a=\kappa _{\text{ }b}^aV^b.  \label{c}
\end{equation}

The space-time metric is postulated to be
\begin{equation}
g_{\mu \nu }=\eta _{ab}V_\mu ^aV_\nu ^b
\end{equation}
with $\eta _{ab}=\left( -1,1,1,1\right) $. Thus the corresponding curvature
is given by (\ref{och}), but now (\ref{nuev}) does not correspond to the
space-time torsion because the vierbein is not given by $e^a$. The
space-time torsion ${\cal T}$ $^a$ is given by
\begin{equation}
{\cal T}\ ^a=DV^a=T^a+R^{ab}\xi _b.
\end{equation}

\subsubsection{\bf Hilbert action invariant under the Poincar\`e group}

Within the $\left( SWGN\right) $-formalism the Hilbert action can be
rewritten as
\begin{equation}
S_{EH}=\int \varepsilon _{abcd}V^aV^cR^{cd}
\end{equation}
which is invariant under general coordinate transformations, under local
Lorentz rotations, as well as under local Poincare translations. In fact
\begin{equation}
\delta S_{EH}=\int \varepsilon _{abcd}\delta \left( R^{ab}V^cV^d\right)
\end{equation}

\begin{equation}
\delta S_{EH}=2\int \varepsilon _{abcd}R^{ab}V^c\delta V^d=0.
\end{equation}
Thus the action is genuinely invariant under the Poincar\`e group
{\bf without\ imposing a torsion-free condition.}

The variations of the action with respect to $\xi ^a,$ $e^a,$ $\omega ^{ab}$
lead to the following equations:
\begin{equation}
\varepsilon _{abcd}{\cal T}\ ^bR^{cd}=0
\end{equation}

\begin{equation}
\varepsilon _{abcd}V^bR^{cd}=0
\end{equation}

\begin{equation}
\ \ \ \ \varepsilon _{[aed}\xi _{b]}V^eR^{cd}+\varepsilon
_{abcd}V^c{\cal T} ^d=0
\end{equation}

which reproduce the correct Einstein equations \cite{grigna}:
\begin{equation}
{\cal T}\ ^a=DV^a=0  \label{tors}
\end{equation}
\begin{equation}
\varepsilon _{abcd}V^bR^{cd}=0.
\end{equation}
The commutator of two local Poincar\`e translations is given by
\begin{equation}
\left[ \delta (\rho _2),\delta (\rho _1)\right] =0
\end{equation}
i.e. the local Poincar\`e translations now commute. The rest of the algebra
is unchanged. Thus the Poincar\`e algebra closes off-shell. This fact has
deep consequences in supergravity \cite{sal}.

\subsection{\bf Lanczos-Lovelock Theory invariant under the Poincar\`e group}

\subsubsection{{\bf Invariance of }$LL${\bf \ action}}

Within the $(SWGN)$-formalism the action for $LL$ gravity can be rewritten
as
\[
S=\int \sum_{p=0}^{\left[ d/2\right] }\alpha _p\varepsilon _{a_1a_2\cdot
\cdot \cdot \cdot \cdot a_d}R^{a_1a_2}\cdot \cdot \cdot
\]
\begin{equation}
\cdot \cdot \cdot R^{a_{2p-1}a_{2p}}V^{a_{2p+1}}\cdot \cdot \cdot \cdot
V^{a_d}  \label{quince}
\end{equation}
with $R^{ab}=d\omega ^{ab}+\omega _c^a\omega ^{cb}$ and $V^a$ is
given in (\ref{catorce}). This action is, both when odd dimensions
and when even dimensions are considered, invariant under local
Lorentz rotations and under local Poincar\`e translations as well
as under diffeomorphism. In fact
\[
\delta S=\int \sum_{p=0}^{\left[ d/2\right] }\alpha _p\varepsilon
_{a_1a_2\cdot \cdot \cdot \cdot \cdot \cdot a_d}R^{a_1a_2}\cdot \cdot \cdot
\]
\begin{equation}
\cdot \cdot \cdot R^{a_{2p-1}a_{2p}}V^{a_{2p+1}}\cdot \cdot \cdot \cdot
\delta V^{a_d}=0
\end{equation}
where we have used the invariance of $V^a$ under local Poincar\'e
translation (\ref{poinc}). Thus the $LL$ action is genuinely invariant under
the Poincar\'e group.

\subsubsection{\bf Equations of motions}

The variations of the acti\'on with respect to $e^a,\omega ^{ab},\xi ^a$
lead to the following equations of motion:
\[
\sum_{p=0}^{\left[ \left( d-1\right) /2\right] }\alpha _p(d-2p)\varepsilon
_{a_1a_2\cdot \cdot \cdot \cdot \cdot \cdot a_d}R^{a_1a_2}\cdot \cdot \cdot
\cdot \cdot
\]
\begin{equation}
\cdot \cdot \cdot \cdot \cdot R^{a_{2p-1}a_{2p}}V^{a_{2p+1}}\cdot \cdot
\cdot \cdot V^{a_{d-1}}=0  \label{I}
\end{equation}
\[
\sum_{p=1}^{\left[ \left( d-1\right) /2\right] }p(d-2p)\alpha _p\varepsilon
_{aba_3\cdot \cdot \cdot \cdot \cdot \cdot a_d}R^{a_3a_4}\cdot \cdot
R^{a_{2p-1}a_{2p}}{\cal T}^{a_{2p+1}}\bullet
\]
\[
\bullet V^{a_{2p+2}}\cdot \cdot \cdot \cdot V^{a_d}+\left( -1\right)
^{d-1}\sum_{p=0}^{\left[ \left( d-1\right) /2\right] }(d-2p)\alpha _p\bullet
\]
\begin{equation}
\bullet \varepsilon _{a_1\cdot \cdot \cdot \cdot \cdot
a_{d-1}}R^{a_1a_2}\cdot \cdot \cdot R^{a_{2p-1}a_{2p}}V^{a_{2p+1}}\cdot
\cdot \cdot V^{a_{d-1}}\xi _b=0  \label{II}
\end{equation}
\[
\sum_{p=0}^{\left[ \left( d-1\right) /2\right] }\alpha
_p(d-2p)(d-2p-1)\varepsilon _{a_1a_2\cdot \cdot \cdot \cdot \cdot \cdot
a_d}R^{a_1a_2}\cdot \cdot \cdot
\]
\begin{equation}
\cdot \cdot \cdot R^{a_{2p-1}a_{2p}}V^{a_{2p+1}}\cdot \cdot \cdot \cdot
V^{a_{d-2}}{\cal T}^{a_{d-1}}=0.  \label{III}
\end{equation}

The field equation corresponding to the variation of the action
(\ref{quince}) with respect to $\xi ^a$ is not an independent
equation. In fact, taking the covariant derivative operator $D$ of
equation (\ref{I}) we obtain the same equation that one obtains by
varying the action (\ref{quince}) with respect to $\xi ^a.$
Furthemore, equations (\ref{I}),(\ref{II}) are also not completely
independent. In fact taking the product of (\ref{I}) with $-\xi _b
$ and then taking the addition with (\ref{II}) we obtain
\[
\sum_{p=1}^{\left[ \left( d-1\right) /2\right] }p(d-2p)\alpha _p\varepsilon
_{aba_3\cdot \cdot \cdot \cdot \cdot \cdot a_d}R^{a_3a_4}\cdot \cdot \cdot
\cdot \cdot
\]
\begin{equation}
\cdot \cdot \cdot \cdot \cdot R^{a_{2p-1}a_{2p}}{\cal T}
^{a_{2p+1}}V^{a_{2p+2}}\cdot \cdot \cdot \cdot V^{a_d}=0.
\label{IV}
\end{equation}
This means that the equations (\ref{I}),(\ref{IV}) are
independent. The equations (\ref{I}),(\ref{IV}) can be rewritten
in the the form (\ref{tres} ), (\ref{cuatro}), where now
\[
\varepsilon _a^p:=\varepsilon _{ab_1b_2\cdot \cdot \cdot \cdot \cdot \cdot
b_{d-1}}R^{b_1b_2}\cdot \cdot \cdot \cdot
\]
\begin{equation}
\cdot \cdot \cdot \cdot R^{b_{2p-1}b_{2p}}V^{b_{2p+1}}\cdot \cdot \cdot
V^{b_{d-1}}  \label{V}
\end{equation}
\[
\varepsilon _{ab}^p=\varepsilon _{aba_3\cdot \cdot \cdot \cdot \cdot \cdot
a_d}R^{a_3a_4}\cdot \cdot \cdot \cdot
\]
\begin{equation}
\cdot \cdot \cdot \cdot R^{a_{2p-1}a_{2p}}T^{a_{2p+1}}V^{a_{2p+2}}\cdot
\cdot \cdot V^{a_d}  \label{VI}
\end{equation}
with ${\cal T}^a=DV^a$.

It is easy to check that, taking the covariant derivative of (\ref{I}) and
using the Bianchi identities, one obtains
\begin{equation}
D\varepsilon _a^p=(d-2p-1)V^b\varepsilon _{ba}^{p+1},  \label{VII}
\end{equation}
for $0\leq p\leq \left[ \left( d-1\right) /2\right] ,$ which leads to the
following off-shell identity:
\begin{equation}
D\varepsilon _a=\sum_{p=1}^{\left[ \left( d+1\right) /2\right] }\alpha
_{p-1}(d-2p+2)(d-2p+1)V^b\varepsilon _{ba}^p  \label{VIII}
\end{equation}
which, by consistency with (\ref{tres}), must also vanish. On the other
hand, taking the exterior product of (\ref{cuatro}) with $V^b$ one obtains
\begin{equation}
V^b\varepsilon _{ba}=\sum_{p=1}^{\left[ \left( d-1\right) /2\right] }\alpha
_pp(d-2p)V^b\varepsilon _{ba}^p  \label{IX}
\end{equation}
which, by consistency with (\ref{cuatro}), must also vanish.

Now we show that, following the same method used in
ref.\cite{tron}, the action (\ref{quince}) has the same
coefficients as those obtained in ref. \cite{tron} for the action
(\ref{uno}).

\subsubsection{{\bf Coefficients }$\alpha _p${\bf \ for }$d=2n-1$}

Following the same procedure of ref. \cite{tron} one can see that
in odd dimensions, (\ref{VIII}) and (\ref{IX}) have the same
number of terms because the last term in (\ref{VIII}) vanishes.
Now, if (\ref{VIII}) and (\ref{IX}) are to impose no further
algebraic constraint on $R^{ab}$ and $T^a$ , then $D\varepsilon
_a$ and $V^b\varepsilon _{ba}$ must be proportional term by term,
which implies the following recursion relation for the
coefficients:
\begin{equation}
\gamma \frac{\alpha _p}{\alpha _{p+1}}=\frac{(p+1)(d-2p+2)}{(d-2p)(d-2p+1)}
\end{equation}
whose solution is given by (\ref{nueve}). Thus, the action contain only two
fundamental constants $\alpha _0$ and $\gamma ,$ related to the
gravitational and the cosmological constant.

Therefore, the use of the $(SWGN)$-formalism does not change the
coefficients $\alpha _p$ of the action already obtained in ref.\cite{tron},
but now the $LL$ action for $d=2n-1$ is genuinelly invariant under the
Poincar\'{e} group.

 >From the action (\ref{quince}) and the equations (\ref{I}),
(\ref{II}), (\ref {III}) we can see that, once the gauge $\xi
^a=0$ is chosen, from equations ( \ref{catorce}), (\ref{tors}) it
follows that $V^a=e^a,$ ${\cal T}\ ^a=T\ ^a=De^a,$ and that the
action (\ref{quince}) takes the form (\ref{uno}) and the equations
(\ref{I}), (\ref{II}), (\ref{III}) take the forms of the equations
for $LL$ gravity theory as developed in refs. \cite{teit}, \cite
{tron}$.$

We must also note that the action (\ref{uno}) for $d=2n-1$ is invariant
under the Poincar\'e group only when $\alpha _p=\alpha _n=1$ and $\alpha
_p=0 $ $\forall $ $p\neq n.$ In this case both formalisms coincide. In fact,
for $\alpha _n=1$ and $\alpha _p=0,$ (\ref{quince}) takes the form
\begin{equation}
S=\int \varepsilon _{a_1a_2\cdot \cdot \cdot \cdot \cdot \cdot
a_d}R^{a_1a_2}\cdot \cdot \cdot \cdot R^{a_{d-2}a_{d-1}}V^{a_d}
\end{equation}
with $V^d=e^d+D\xi ^d$. Using the Bianchi identities it is direct to see
that, up to a surface term,
\begin{equation}
S=\int \varepsilon _{a_1a_2\cdot \cdot \cdot \cdot \cdot \cdot
a_d}R^{a_1a_2}\cdot \cdot \cdot \cdot R^{a_{d-2}a_{d-1}}e^{a_d}.
\end{equation}

\subsubsection{{\bf Coefficients }$\alpha _p${\bf \ for }$d=2n$}

In even dimensions, equation (\ref{VIII}) has one more term than (\ref{IX}).
This means that (\ref{VIII}) and (\ref{IX}) cannot be compared term by term.
Following the same procedure of ref. \cite{tron} we find that for $d=2n$

\begin{equation}
\varepsilon _{ab}=D{\cal \chi }_{ab}  \label{tau}
\end{equation}
where
\begin{equation}
{\cal \chi }_{ab}=\sum_{p=1}^{\left[ \left( d-1\right) /2\right]
}p\alpha _p {\cal \chi }_{ab}^p
\end{equation}
and
\[
{\cal \chi }_{ab}^p=\varepsilon _{aba_3a_4\cdot \cdot \cdot \cdot \cdot
\cdot \cdot a_d}R^{a_3a_4}\cdot \cdot \cdot
\]
\begin{equation}
R^{a_{2p-1}a_{2p}}V^{a_{2p+1}}\cdot \cdot \cdot V^{a_d}
\end{equation}
\begin{equation}
V^b{\cal \chi }_{ba}^p=\varepsilon _a^{p-1}  \label{tau1}
\end{equation}
\begin{equation}
D{\cal \chi }_{ab}^p=(d-2p)\varepsilon _{ab}^p.  \label{tau2}
\end{equation}

 >From (\ref{tau1}) and (\ref{tau2}) we find that (\ref{VIII}) can also be
written

\[
D\varepsilon _a=\sum_{p=1}^{\left[ n-1\right] }2\alpha
_{p-1}(n-p+1){\cal \chi }_{ab}^p{\cal T}^b
\]
\begin{equation}
-\sum_{p=1}^{\left[ n-1\right] }4\alpha _{p-1}(n-p+1)(n-p)V^b\varepsilon
_{ba}^p.  \label{X}
\end{equation}

This equation can be compared with (see \ref{IX})
\begin{equation}
V^b\varepsilon _{ba}=\sum_{p=1}^{n-1}2p\alpha _p(n-p)V^b\varepsilon _{ba}^p.
\label{XI}
\end{equation}

Both (\ref{tau}) and (\ref{XI}) can be zero if either ${\cal
T}^a=0,$ or $ {\cal \chi }_{ab}=0.$ These conditions are excessive
for the vanishing of ( \ref{X}). In the same way as in ref.
\cite{tron}, it is sufficient to impose the weaker conditions
${\cal T}^a{\cal \chi }_{ab}=0,$ and, at the same time, that the
second term in (\ref{X}) be proportional to (\ref{XI}). Now, (
\ref{X}) and (\ref{XI}) possess the same number of terms, which
implies the following recursion relation for the coefficients:
\begin{equation}
2\gamma (n-p+1)\alpha _{p-1}=p\alpha _p,  \label{XII}
\end{equation}
whose solution is given by (\ref{trece}). Thus, the action for $d=2n$
contains only two fundamental constants $\alpha _0$ and $\gamma ,$ related
to the gravitational and the cosmological constants.

Therefore, in the same way as for $d=2n-1,$ the use of the
$(SWGN)$ -formalism does not change the coefficients $\alpha _p$
of the action already obtained in ref.\cite{tron}. However, now
the $LL$ action for $d=2n$ is genuinely invariant under the
Poincar\'{e} group.

 >From the action (\ref{quince}) and the equations (\ref{I}),
(\ref{II}), (\ref {III}) we can see that, once the gauge $\xi
^a=0$ is chosen, from equations ( \ref{catorce}), (\ref{tors}) it
follows that $V^a=e^a,$ ${\cal T}\ ^a=T\ ^a=De^a,$ and that the
action (\ref{quince}) takes the form (\ref{uno}) and the equations
(\ref{I}), (\ref{II}), (\ref{III}) take the forms of the equations
for $LL$ gravity theory as developed in refs. \cite{teit}, \cite
{tron}$.$

We must to note that, within the $SWGN$- formalism, the action (\ref{d1})
can be rewritten as
\[
S=\sum \frac \kappa {d-2p} {k \choose p} l^{2(p-k)}\varepsilon
_{a_1a_2\cdot \cdot \cdot \cdot \cdot \cdot a_d}R^{a_1a_2}\cdot
\cdot \cdot
\]
\begin{equation}
\cdot \cdot \cdot \cdot R^{a_{2p-1}a_{2p}}V^{a_{2p+1}}\cdot \cdot \cdot
\cdot V^{a_d}  \label{d7}
\end{equation}
which is invariant under local Lorentz rotations as well as under Poincar\'e
translations.

\section{Comments}

We have shown in this work that the successful formalism used by Stelle-West
\cite{stelle} and Grignani-Nardelli \cite{grigna} to construct an action for
$\left( 3+1\right) $-dimensional gravity invariant under the Poincar\`{e}
group can be generalized to arbitrary dimensions. The main result of this
paper is that we have shown that the $\left( SWGN\right) $ formalism
permits, both when odd dimensions and when even dimensions are considered,
constructing a higher dimensional gravity action invariant under local
Lorentz rotations and under local Poincare translations. This means using
the vierbein $V^a$ which involves in its definition the so called ''Poincare
coordinates'' $\xi ^a(x)$. It is also proved that such actions have the same
coefficients as those obtained in ref. \cite{tron}.

It is perhaps interesting to note that if one considers, following ref. \cite
{grigna}$,$ $g_{\mu \nu }=V_\mu ^aV_\nu ^b\eta _{ab}$, one can write the
lagrangian of the action (\ref{quince}) in the form that was written in ref.
\cite{teit}. This means that, if one considers the theory constructed in
terms of the space-time metric $g_{\mu \nu }$, ignoring the underlying
formulation, the theory described in our manuscript is completely equivalent
to the theory developed in refs. \cite{teit}, \cite{lovel}. No trace of the
new structure of the vierbein existing in the underlying formulation of the
theory can be found at the metric level.

{\bf Acknowlegments}

This work was supported in part by FONDECYT through Grants \# 1010485 (M.C
andP.S) and \# 1000305 (S del C) and in part by UCV through Grant UCV-DGIP
\# 123.752/00.

\end{document}